\documentclass[12pt]{iopart}

\usepackage{graphicx,xcolor}
\usepackage{cite}
\usepackage{marginnote}
\expandafter\let\csname equation*\endcsname\relax
\expandafter\let\csname endequation*\endcsname\relax
\usepackage{amsmath, amsthm, amssymb}
\UseRawInputEncoding

\newcommand{\ie}{{\emph{i.e.}}}
\usepackage{hyperref}
\hypersetup{colorlinks=true,linkcolor=black}
\begin{document}
\title{Effects of reversed magnetic shear on the plasma rotation stabilization of resistive wall modes in tokamaks}
	
	\author{Sui Wan}
	
	\address{State Key Laboratory of Advanced Electromagnetic Technology, \\International Joint Research Laboratory of Magnetic Confinement Fusion and Plasma Physics, School of Electrical and Electronic Engineering,
		\\	Huazhong University of Science and Technology, Wuhan, 430074,
		China}
	
	\author{Ping Zhu*}
	
	\address{State Key Laboratory of Advanced Electromagnetic Technology, \\International Joint Research Laboratory of Magnetic Confinement Fusion and Plasma Physics, School of Electrical and Electronic Engineering,
		\\	Huazhong University of Science and Technology, Wuhan, 430074,
		China;
		~\\
		Department of Nuclear Engineering and Engineering Physics, 
		\\University of Wisconsin-Madison, Madison,
		Wisconsin, 53706, United States of America}
	\ead{zhup@hust.edu.cn}
	
	\author{Linjin Zheng}
	
	\address{Institute of Fusion Studies, University of Texas at Austin, Austin, Texas, 78712, United States of America}
	\vspace{10pt}
	\begin{indented}
		\item[] \today
	\end{indented}
	\begin{abstract}
		Effects of reversed magnetic shear on the plasma rotation stabilization of resistive wall modes in tokamaks are investigated using the AEGIS code. MHD equilibria in toroidal configuration from circular cross-sections to realistic CFETR-like scenarios with various magnetic shear profiles are considered. Two critical aspects of the $n=1$ RWM are examined: the influence of toroidal rotation on the unstable regime and the toroidal rotation frequency thresholds required for complete stabilization. It is found that strongly reversed magnetic shear consistently broadens the unstable $\beta_{\rm N}$ window in both circular and CFETR equilibria when toroidal rotation is included.
		 Furthermore, reversed magnetic shear significantly reduces the rotational stabilization, resulting in narrower stability windows and notably higher toroidal rotation frequency thresholds required for complete RWM suppression compared to the cases with positive shear only. These results clearly demonstrate that the reversed magnetic shear in the advanced tokamak configuration imposes more stringent requirement for the effective toroidal rotation stabilization of the $n=1$ RWM.
	\end{abstract}
	
	\vspace{2pc}
	
	\section{Introduction}\label{sec1}
	
	High normalized beta, defined as $\beta_{\mathrm{N}}=\beta[\%]a[\mathrm {m}]B_t[\mathrm{T}]/I_p[\mathrm{MA}]$, is essential for fusion performance in tokamaks, where $\beta$ is the ratio of the volume-averaged plasma pressure to magnetic pressure, $a$ the minor radius, $B_t$ the vacuum toroidal magnetic field, and $I_p$ the total plasma current. However, the achievable $\beta_{\mathrm{N}}$ is limited by magnetohydrodynamic (MHD) instabilities, particularly the external kink modes~\cite{troyon84a}, which can become unstable above the no-wall limit ($\beta_{\mathrm{N}}^{\mathrm{no\mbox{-}wall}}$). Although a close perfect conducting wall can fully suppress external kink modes up to the higher ideal-wall limit ($\beta_{\mathrm{N}}^{\mathrm{ideal\mbox{-}wall}}$), any realistic wall with finite resistivity allows a residual kink - resistive wall mode (RWM) to grow on a slowed timescale between these limits, \ie~$\beta_{\mathrm{N}}^{\mathrm{no\mbox{-}wall}}<\beta_{\mathrm{N}}<\beta_{\mathrm{N}}^{\mathrm{ideal\mbox{-}wall}}$~\cite{joffrin07a,chu10a}. Although RWMs grow on a relatively slow resistive timescale, they are one of the leading causes to disruptions, thus posing a critical obstacle to achieving steady-state, high-$\beta_{\mathrm{N}}$ operation in next-generation fusion devices such as International Thermonuclear Experimental Reactor (ITER)~\cite{gormezano07a} and the China Fusion Engineering Test Reactor (CFETR)~\cite{wan14a,zhuang19a}.
	
	Both active and passive control techniques have been extensively explored to stabilize RWMs in the reversed field pinch (RFP) and tokamaks. Active methods typically involve external coils designed to detect plasma perturbations and apply magnetic feedback, enabling direct and real-time suppression of RWMs~\cite{strait99a,brunsell05a,liu06a,garofalo07a,Katsuro-Hopkins07a,zanca12a,park14a,piovesan14a,setiadi16a,ren18a,clement18a,xia19b,pigatto19a,wang21a,saad22a,battey23a,xia23a,saad25a}. Although active control has demonstrated clear benefits in experiments, the associated engineering complexity and additional power demands make its implementation challenging in large-scale fusion devices. In contrast, passive control techniques rely on intrinsic damping mechanisms, including interactions between plasma and surrounding conducting structures, as well as stabilizing effects related to plasma rotation~\cite{bondeson94a,ward95a,chu95a,betti95a,zheng05a,takechi07a,han20a,han23a,zheng24a}, flow shear~\cite{chu95a,liu15a}, and particle resonances~\cite{hu04a,liu09a,liu10a,berkery10a,zheng10pop,hao11a,wang12a,menard14a}. Eddy current dissipation in the resistive wall slows the mode growth rate, while internal damping mechanisms, such as Alfv\'{e}n and acoustic continuum damping, drift-kinetic resonance with thermal and fast ions, are capable of substantially raising the stability threshold. Thus, achieving effective passive stabilization through optimized plasma parameters remains an indispensable scheme for future reactor-scale operations.
	
	In advanced tokamak (AT) scenarios, optimization of the plasma current profile, primarily by increasing the fraction of bootstrap current and reducing the need for auxiliary current drive --- naturally leads to reversed magnetic shear in the core region. Reversed shear equilibria often feature steep pressure gradients and significant off-axis bootstrap currents, making them more susceptible to RWMs~\cite{hender07a}. Additionally, previous studies indicate that reversed magnetic shear in the plasma core can enhance the instability of low-$n$ (where $n$ is the toroidal mode number) modes~\cite{huysmans99a,takeji02a,turnbull02a,wan24a}. In the absence of rotation and kinetic effects, reversed shear configurations typically show a higher $\beta_{\mathrm{N}}^{\mathrm{ideal-wall}}$ and a lower $\beta_{\mathrm{N}}^{\mathrm{no-wall}}$ for low-$n$ modes in comparison to the positive shear cases, resulting in a broader unstable $\beta_{\rm N}$ window for RWMs~\cite{manickam94a,zheng10pop}. Whereas the toroidal rotation is known to affect both the ideal-wall and no-wall $\beta_{\mathrm{N}}$ limits~\cite{ward95a,menard14a}, how such rotational effects on RWMs may vary in reversed shear configurations remains unclear. Previous works have compared weakly and strongly reversed shear equilibria~\cite{zheng10pop}, primarily in the context of ITER scenarios, demonstrating that stronger reversed shear generally requires higher rotation frequencies for complete RWM suppression. However, other numerical studies suggest that strong toroidal rotation may re-destabilize RWMs through coupling with stable ideal MHD eigenmodes, particularly when the Doppler-shifted frequency aligns with that of a stable mode~\cite{aiba13a,aiba15a,aiba15b}. More importantly, it is desirable to obtain a more complete perspective on the effects of toroidal rotation on RWM stability across a full range of magnetic shear conditions - from positive to strongly reversed shear - in both circularly and non-circularly shaped equilibria.
	
	In this study, the stability of RWMs in configurations with positive and reversed magnetic shear is investigated for both circular and non-circular plasma cross-sections. The AEGIS eigenvalue code~\cite{zheng06a} is employed to evaluate the effects of uniform toroidal rotation on the ideal-wall and no-wall $\beta_{\mathrm{N}}$ limit, as well as the rotation frequency thresholds required for complete RWM stabilization. It is found that strongly reversed shear configurations tend to expand the unstable RWM regime and demand higher rotation frequency for stabilization.
	
	The remainder of this paper is organized as follows. Section~\ref{sec2} describes the equilibrium configurations employed in this work. Section~\ref{sec3} outlines the numerical model and methods used in the AEGIS code. Section~\ref{sec4} presents the numerical results and analysis. Finally, Section~\ref{sec5} summarizes the main conclusions and discusses their implications.
	\section{Equilibrium}\label{sec2}
	
\begin{figure}[h]
		\centering
		\includegraphics[width=0.8\linewidth]{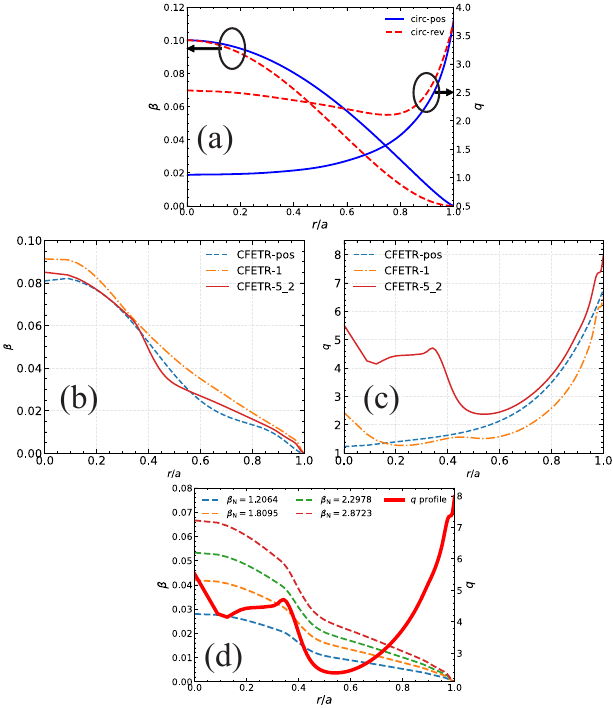}
		\caption{(a) Equilibrium pressure ($\beta=2\mu_0p/B^2_0$) and safety factor profiles for circular cross-section equilibria: \texttt{circ-pos} (solid blue line, positive magnetic shear) and \texttt{circ-rev} (dashed red line, reversed magnetic shear). (b) Equilibrium $\beta$ and (c) safety factor profiles for CFETR scenarios: \texttt{CFETR-5\_2} (solid red line, strongly reversed shear), \texttt{CFETR-1} (dash-dotted orange line, weak reversed shear), and \texttt{CFETR-pos} (dashed blue line, monotonic safety factor). (d) Pressure profiles ($\beta$, dashed lines) for various $\beta_{\rm N}$ values based on the \texttt{CFETR-5\_2} equilibrium, for the same safety factor profile (solid red line).}
		\label{fig:eq}
	\end{figure}

	In this work, plasma equilibria with two distinct cross-sections are generated using equilibrium codes CHEASE~\cite{lutjens96a} and EFIT~\cite{lao85a}. The first type, computed using CHEASE, features a circular cross-section with pressure and safety factor profiles corresponding to positive and reversed magnetic shear equilibria previously studied in~\cite{kessel94a,manickam94a}, which are denoted as \texttt{circ-pos} and \texttt{circ-rev} here. The main parameters for these equilibria include an aspect ratio of $R_0/a=4.5$ and a toroidal magnetic field at magnetic axis $B_t=1$ T. The resulting equilibrium profiles are shown in Fig.~\ref{fig:eq}(a). The second type corresponds to the CFETR-like equilibria based on two baseline steady-state scenarios~\cite{zhuang19a,chen21a,chen21b,zhou22a}. The first scenario, named \texttt{CFETR-5\_2}, has a major radius $R_0=7.2$ m, minor radius $a=2.2$ m, $B_t=6.53$ T and the total plasma current $I_p=11$ MA. It features a strongly negative magnetic shear in the core region. The second scenario, named \texttt{CFETR-1}, shares the same geometry and magnetic field parameters with \texttt{CFETR-5\_2} but operates at a higher total plasma current of $I_p=13$ MA, and has a weaker negative magnetic shear in the core. Additionally, a monotonic safety factor profile is designed using EFIT code for comparison, referred to as \texttt{CFETR-pos}. The equilibrium profiles for CFETR scenarios are presented in Fig.~\ref{fig:eq}(b)(c). The key parameters for all equilibrium configurations are summarized in Table~\ref{tab:eq_params}.
	
	\begin{table}[htbp]
		\centering
		\caption{Summary of equilibrium parameters for different configurations.}
		\begin{tabular}{lcccccc}
			\hline
			Equilibrium  & $R_0$ (m) & $a$ (m) & $B_t$ (T)  & Magnetic shear in the core region \\ \hline
			\texttt{circ-pos}  & 4.5 & 1.0 & 1.00  & positive \\
			\texttt{circ-rev}  & 4.5 & 1.0 & 1.00  & reversed \\
			\texttt{CFETR-5\_2}  & 7.2 & 2.2 & 6.53  & strongly reversed \\
			\texttt{CFETR-1}  & 7.2 & 2.2 & 6.53 &  weak reversed \\
			\texttt{CFETR-pos}  & 7.2 & 2.2 & 6.53  & positive \\ \hline
		\end{tabular}
		\label{tab:eq_params}
	\end{table}

	To investigate the $\beta_{\rm N}$ limit, a series of plasma equilibria with different $\beta_{\rm N}$ values are constructed by varying the core pressure $\beta_0$ while keeping the safety factor profile fixed. These equilibria are used to determine the ideal-wall and no-wall $\beta_{\mathrm{N}}$ limits. For example, Fig.~\ref{fig:eq}(d) presents the equilibrium profiles based on the \texttt{CFETR-5\_2} configuration, showing pressure profiles with various $\beta_{\rm N}$ values.
 
	\section{Numerical model in AEGIS code}\label{sec3}
	
	In this work, the AEGIS code~\cite{zheng05a,zheng06a,zheng17a} is employed to solve the following linearized ideal MHD equations in the plasma region
	\begin{equation}\label{eq:aegis-equation}
		-\rho_m\hat{\omega}^2\boldsymbol{\xi}=\delta \bi{J}\times\boldsymbol{B}+\boldsymbol{J}\times\delta\boldsymbol{B}-\boldsymbol{\nabla}\delta P,
	\end{equation}
	where $\rho_m$ represents the mass density, $\boldsymbol{\xi}$ is the perpendicular fluid displacement, $\boldsymbol{J}$ and $\boldsymbol{B}$ are the equilibrium current density and magnetic field, respectively. Perturbed quantities satisfy the relationships $\mu_0\delta\boldsymbol{J}=\boldsymbol{\nabla}\times\delta\boldsymbol{B}$, $\delta\boldsymbol{B}=\boldsymbol{\nabla}\times\left(\boldsymbol{\xi}\times\boldsymbol{B} \right) $, and $\delta P=-\boldsymbol{\xi}\cdot \boldsymbol{\nabla}P$. The plasma rotation leads to a Doppler-shifted frequency as $\hat{\omega}=\omega+n\Omega$, where $\Omega$ denotes the toroidal rotation frequency. In tokamak plasmas operating in the subsonic regime, the contributions from centrifugal and Coriolis forces are minimal and can be neglected~\cite{waelbroeck91a,zheng99a}. Furthermore, the influence of plasma compressibility is primarily manifested via the apparent mass effect~\cite{green62a}, which is included in the momentum equation \eqref{eq:aegis-equation} by treating $\rho_m$ as the effective mass. In AEGIS, the vacuum region is modeled using the scalar magnetic potential $u$, which satisfies the Laplace equation $\boldsymbol{\nabla}^2u=0$ and the perturbed magnetic field is determined by $\delta\boldsymbol{B}=-\boldsymbol{\nabla}u$. The solution of Eq.~\eqref{eq:aegis-equation} is then required to match the vacuum and wall responses, thereby forming a complete eigenvalue problem.
	
\begin{figure}[h]
		\centering
		\includegraphics[width=0.6\linewidth]{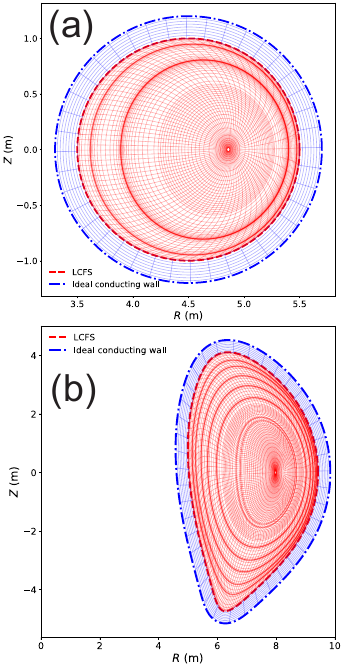}
		\caption{Computational meshes used in AEGIS calculations for equilibria with (a) circular cross-section (\texttt{circ-pos}) and (b) shaped CFETR cross-section (\texttt{CFETR-5\_2}). Both plasma and vacuum regions are included. The conformal conducting wall is located at $r_w/a = 1.2$.}
		\label{fig:aegis-grid}
	\end{figure}

	The AEGIS code employs a radial adaptive shooting method combined with Fourier decomposition in the poloidal and toroidal directions to solve the ideal MHD eigenvalue equations with either ideal conducting or resistive wall conditions. For an ideal conducting wall, the boundary condition $\delta \boldsymbol{B}\cdot\nabla \psi=0$ is imposed at the wall surface. In the case of a resistive wall, a thin-wall approximation is applied to the wall boundary conformal to that of the plasma~\cite{betti95a,zheng06a,freidberg2014ideal}. The computational domain includes the plasma region within the separatrix and the vacuum region extending to the first wall. To avoid numerical singularities in the safety factor at the separatrix, the diverted equilibria are often truncated at 99\% of the normalized poloidal flux $\psi_a$. AEGIS generates adaptive computational grids based on tokamak MHD equilibria, automatically packing radial grids near resonant surfaces to achieve the required resolution. Manual grid packing can also be applied as needed. Figs.~\ref{fig:aegis-grid}(a) and~\ref{fig:aegis-grid}(b) show the AEGIS computational meshes for equilibria with a circular cross-section and the CFETR configuration, respectively. Both plasma and vacuum regions are included, and a conformal shaped wall model is employed in the calculations.
	\section{Numerical results}\label{sec4}
	
	For the equilibria introduced in Section~\ref{sec2}, we use AEGIS to calculate the ideal-wall and no-wall $\beta_{\mathrm{N}}$ limits both with and without uniform toroidal rotation, assuming a fixed wall position at $r_{w}/a=1.2$. Additionally, the critical rotation frequency thresholds required for complete stabilization are evaluated at various wall radii. The dependence of these limits and thresholds on the magnetic shear configurations is the primary focus.

	\subsection{$\beta_{\mathrm{N}}$ limit of static equilibrium}\label{sec4.1}
	
\begin{figure}[h]
		\centering
		\includegraphics[width=0.5\linewidth]{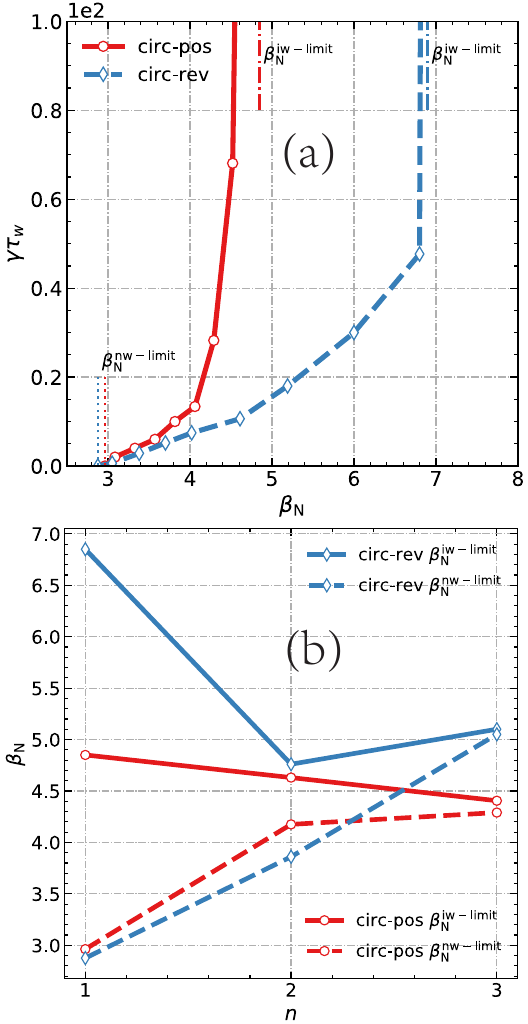}
		\caption{(a) Growth rates of $n=1$ RWM as functions of $\beta_{\rm N}$ for circular cross-section equilibria without rotation. (b) Ideal-wall and no-wall limits for $n=1$-$3$ modes in \texttt{circ-pos} and \texttt{circ-rev} equilibria.}
		\label{fig:circ-beta-limit-no-rot-all}
	\end{figure}

	In the absence of toroidal rotation, the stability boundaries of RWMs are first evaluated for configurations with various magnetic shear introduced in Section~\ref{sec2}. As shown in Fig.~\ref{fig:circ-beta-limit-no-rot-all}(a), the growth rate of $n=1$ RWMs as a function of $\beta_{\rm N}$ for circular cross-section equilibria clearly reveals the no-wall and ideal-wall stability limits: the no-wall limit, defined by the onset of instability ($\gamma\tau_w > 0$), and the ideal-wall limit, characterized by the saturation of growth rates at high $\beta_{\rm N}$. Fig.~\ref{fig:circ-beta-limit-no-rot-all}(b) further presents the extracted $\beta_{\rm N}^{\rm no\mbox{-}wall}$ and $\beta_{\rm N}^{\rm ideal\mbox{-}wall}$ for $n=1\text{--}3$ modes. Compared with the positive shear, the reversed magnetic shear yields higher ideal-wall limits and lower no-wall limits, leading to a wider RWM-unstable window.
	
\begin{figure}[h]
		\centering
		\includegraphics[width=0.7\linewidth]{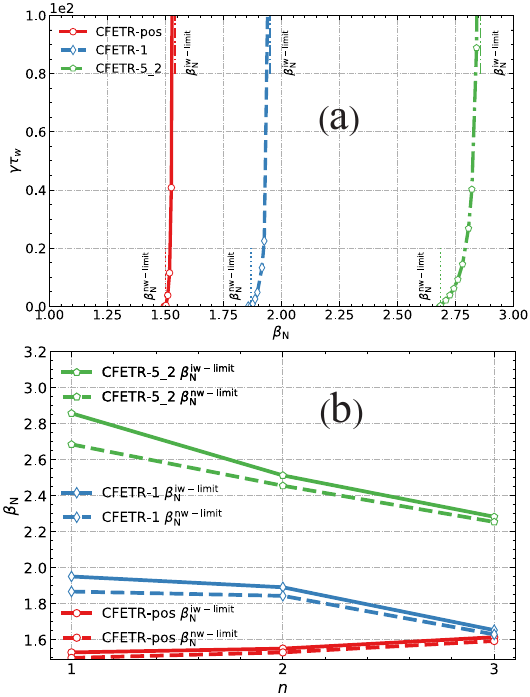}
		\caption{(a) Growth rates of $n=1$ RWM as functions of $\beta_{\rm N}$ for CFETR equilibria without rotation. (b) Ideal-wall and no-wall limits for $n=1$-$3$ modes in \texttt{CFETR-pos}, \texttt{CFETR-1} and \texttt{CFETR-5\_2}.}
		\label{fig:cfetr-betan-limit-no-rot}
	\end{figure}

	For CFETR-based equilibria \texttt{CFETR-5\_2} and \texttt{CFETR-1}, both $\beta_{\rm N}^{\rm ideal-wall}$ and $\beta_{\rm N}^{\rm no-wall}$ decrease with the toroidal mode number $n$ (Fig.~\ref{fig:cfetr-betan-limit-no-rot}). Among these, \texttt{CFETR-5\_2}, with stronger reversed shear, shows the broadest unstable region at low $n$. However, unlike in circular equilibria, reversed shear in CFETR does not always yield higher $\beta_{\rm N}^{\rm ideal-wall}$ or lower $\beta_{\rm N}^{\rm no-wall}$ than its positive shear counterpart, likely due to the off-axis bootstrap current and strong plasma shaping in CFETR which may have weaken the effect of reversed shear on low-$n$ mode stability.
	
\begin{figure}[h]
		\centering
		\includegraphics[width=0.7\linewidth]{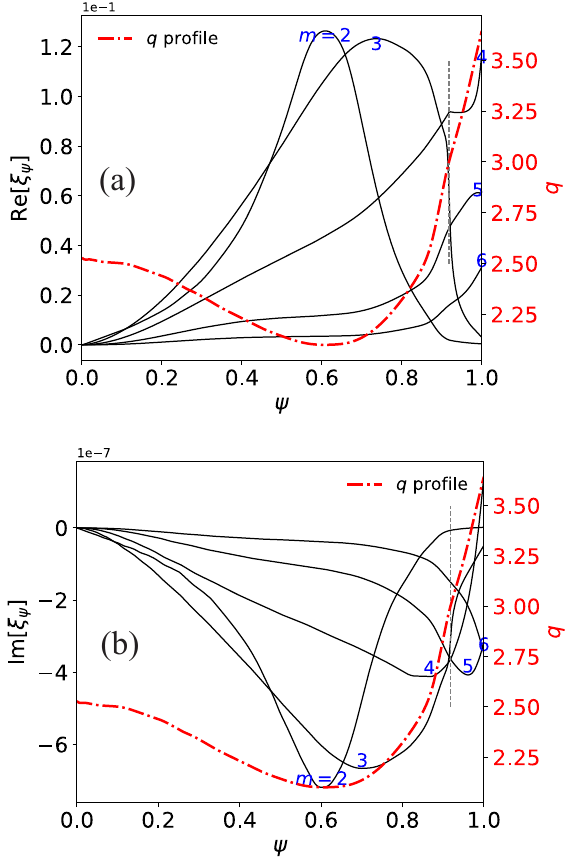}
		\caption{Radial profiles of the real (a) and imaginary (b) parts of the dominant poloidal Fourier components of the radial displacement $\xi_\psi$ for the $n=1$ RWM in the \texttt{circ-rev} equilibrium with $\beta_{\mathrm{N}}=4.91$, $r_w/a=1.2$, in absence of rotation. The dominant poloidal mode numbers are indicated.}
		\label{fig:circ-rev-no-rot-mode}
	\end{figure}

\begin{figure}[h]
		\centering
		\includegraphics[width=1.\linewidth]{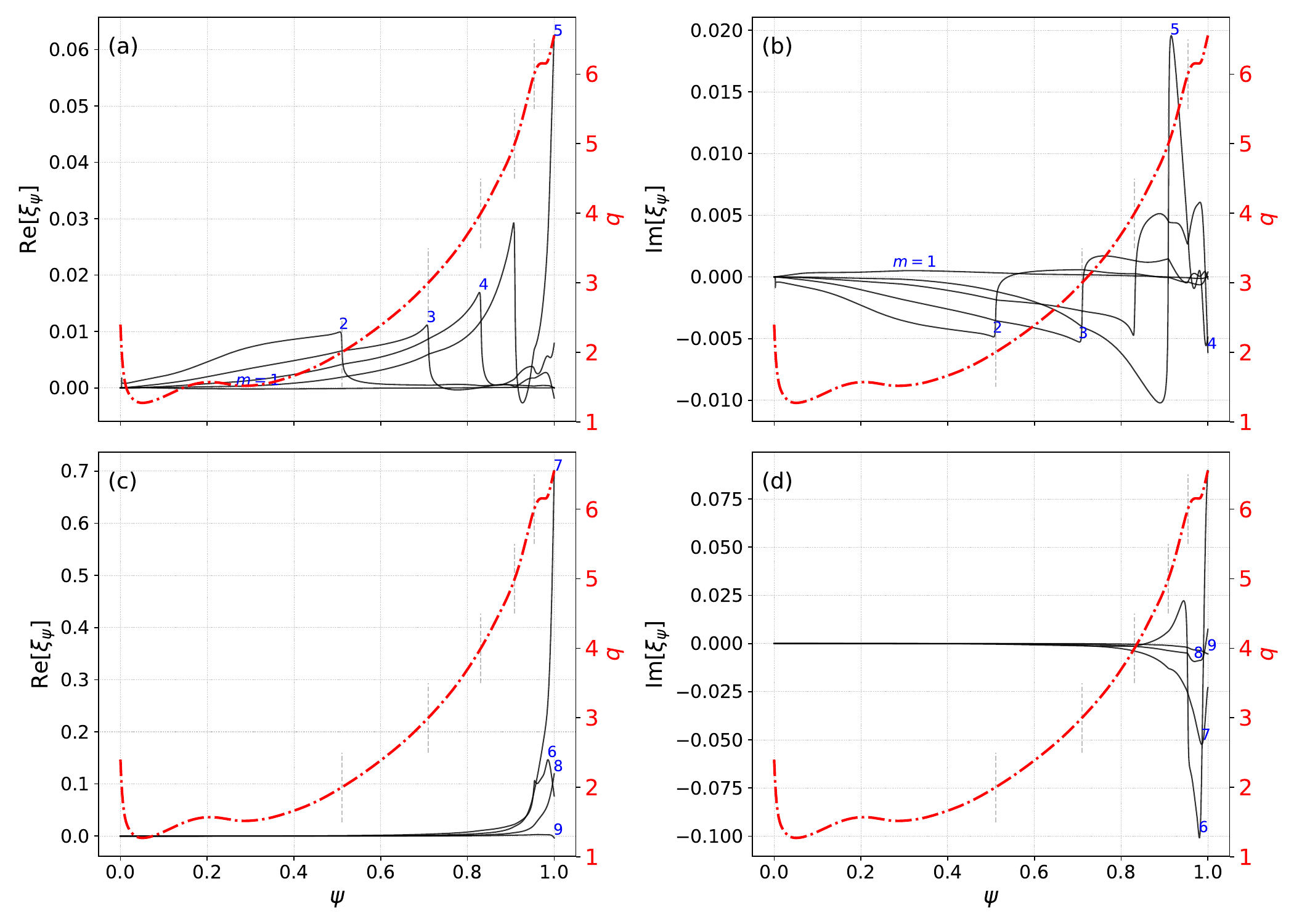}
		\caption{Real (a, c) and imaginary (b, d) parts of the radial displacement $\xi_\psi$ for the $n=1$ RWM in the \texttt{CFETR-1} equilibrium with $\beta_{\mathrm{N}}=1.904$, $r_w/a=1.2$, in absence of rotation. The poloidal harmonics $m=1-5$ (a, b) and $m=6-9$ (c, d) are shown with corresponding labels.}
		\label{fig:cfetr-eq1-no-rot-mode}
	\end{figure}

	The radial displacements $\xi_\psi$ of the $n=1$ RWM, including both real and imaginary parts of its dominant poloidal Fourier components, are shown in Figs.~\ref{fig:circ-rev-no-rot-mode} and~\ref{fig:cfetr-eq1-no-rot-mode} for two cases: \texttt{circ-rev} and \texttt{CFETR-1}, each at a representative $\beta_{\rm N}$ value between the ideal- and no-wall limits. These radial profiles of $\xi_\psi$ reveal the mode structure in terms of both perturbation amplitude and phase variations. For an up-down symmetric circular equilibrium, the imaginary components of $\xi_\psi$ remain negligible, leading to a nearly real eigenmode function in the absence of rotation. In contrast, the \texttt{CFETR-1} equilibrium exhibits sizable imaginary components, even in absence of rotation, due to the up-down asymmetry and 2D shaping of the poloidal cross-section. Similar findings have been also reported in a previous work~\cite{zheng10JCP}.

	\subsection{Effects of toroidal rotation on $\beta_{\mathrm{N}}$ limit}\label{sec4.2}

\begin{figure}[h]
		\centering
		\includegraphics[width=0.7\linewidth]{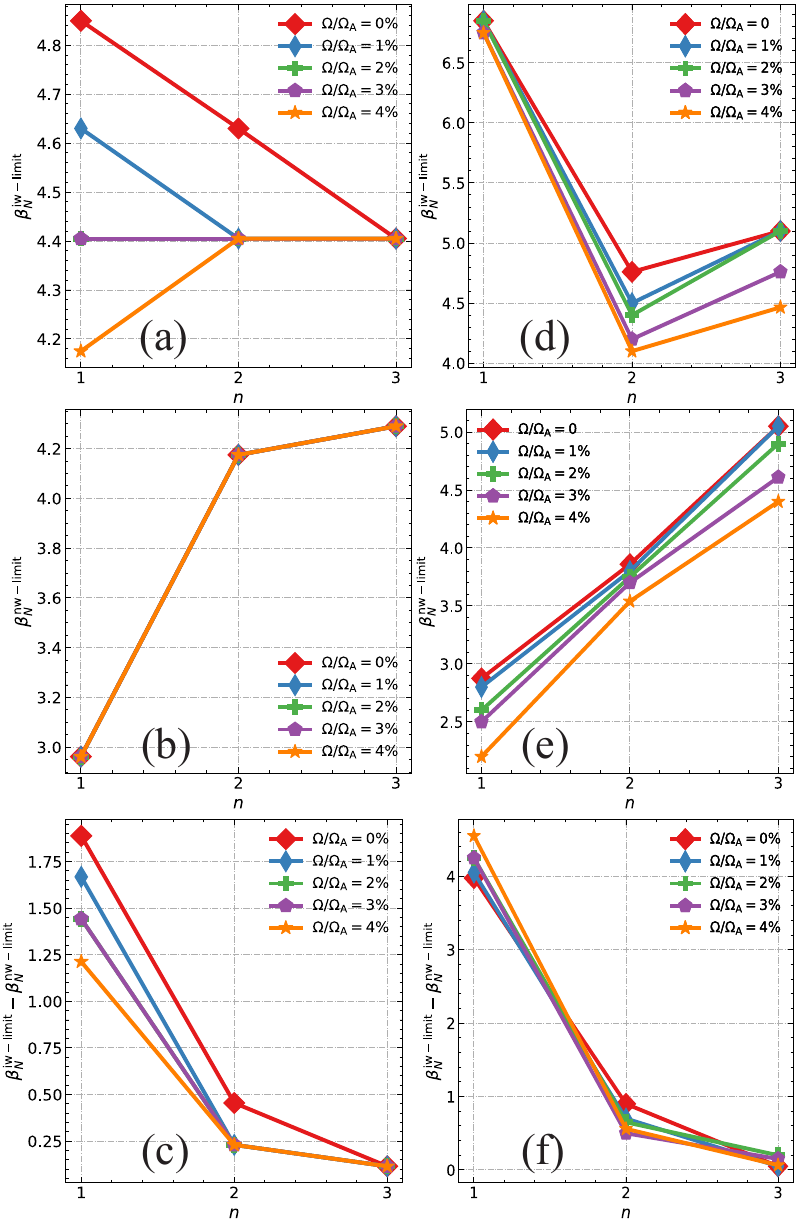}
		\caption{(a) $\beta_{\rm N}^{\rm ideal-wall}$, (b)$\beta_{\rm N}^{\rm no-wall}$ and (c) $\beta_{\rm N}$ limit gap (\ie $\beta_{\rm N}^{\rm ideal-wall}$-$\beta_{\rm N}^{\rm no-wall}$) as functions of toroidal mode number $n$ for \texttt{circ-pos} equilibrium at various toroidal rotation frequencies; (d), (e) and (f) present the same three quantities for the \texttt{circ-rev} equilibrium, respectively.}
		\label{fig:circ-betan-limit-with-rot-all}
	\end{figure}

	For the \texttt{circ-pos} equilibrium, Fig.~\ref{fig:circ-betan-limit-with-rot-all}(a-b) shows that the ideal-wall limit decreases as rotation frequency increases, whereas the no-wall limit remains nearly unchanged, thus the gap $\beta_{\rm N}^{\rm ideal-wall} - \beta_{\rm N}^{\rm no-wall}$ shrinks (Fig.~\ref{fig:circ-betan-limit-with-rot-all}(c)). In contrast, in the \texttt{circ-rev} equilibrium (Fig.~\ref{fig:circ-betan-limit-with-rot-all}(d-f)), both thresholds decrease as rotation frequency increases, but the ideal-wall limit shows a more substantial drop. This results in a widening of the RWM-unstable $\beta_{\mathrm{N}}$-limit gap, in contrast to the positive shear cases. The opposite responses observed in \texttt{circ-pos} and \texttt{circ-rev} cases demonstrate the critical role of magnetic shear in regulating the rotational effects.
	
\begin{figure}[h]
		\centering
		\includegraphics[width=0.4\linewidth]{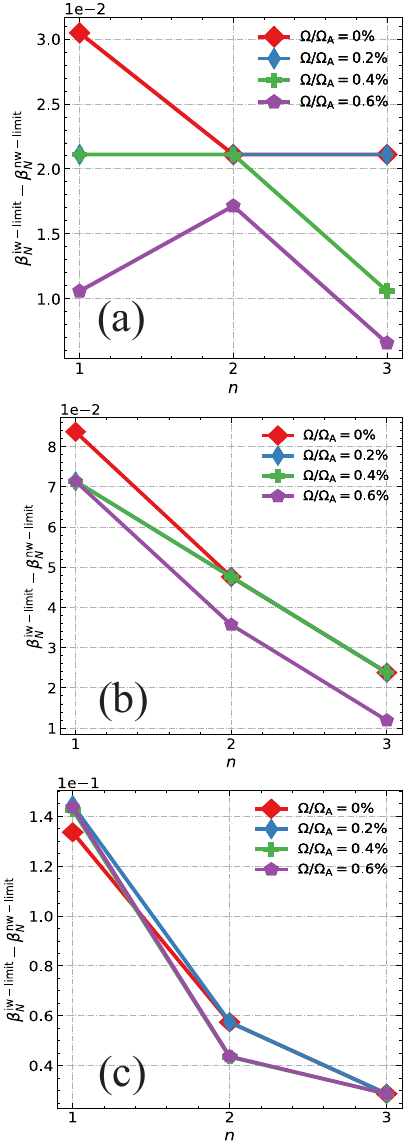}
		\caption{$\beta_{\rm N}$ limit gap ($\beta_{\rm N}^{\rm ideal-wall}$-$\beta_{\rm N}^{\rm no-wall}$) as functions of $n$ for CFETR equilibria: (a) \texttt{CFETR-pos}, (b) \texttt{CFETR-1}, and (c) \texttt{CFETR-5\_2} at various toroidal rotation frequencies.}
		\label{fig:cfetr-betan-limit-with-rot-all}
	\end{figure}

	The same behavior is confirmed in the shaped CFETR equilibria (Fig.~\ref{fig:cfetr-betan-limit-with-rot-all}). In the \texttt{CFETR-pos} and \texttt{CFETR-1} cases, the ideal-wall limit again decreases more rapidly, leading to a narrowing of the unstable gap, similar to the behavior in \texttt{circ-pos}. For the strongly reversed shear case \texttt{CFETR-5\_2}, the unstable window broadens with rotation, consistent with the circular reversed shear case. Notably, the degree of this effect is more pronounced in configurations with stronger negative shear. These results confirm that the influence of toroidal rotation on RWM stability exhibits a strong dependence on the magnetic shear configuration. This dependence will be further examined in Section~\ref{sec4.3}, where the rotation frequency thresholds required for complete RWM suppression are evaluated.
	
\begin{figure}[h]
		\centering
		\includegraphics[width=0.7\linewidth]{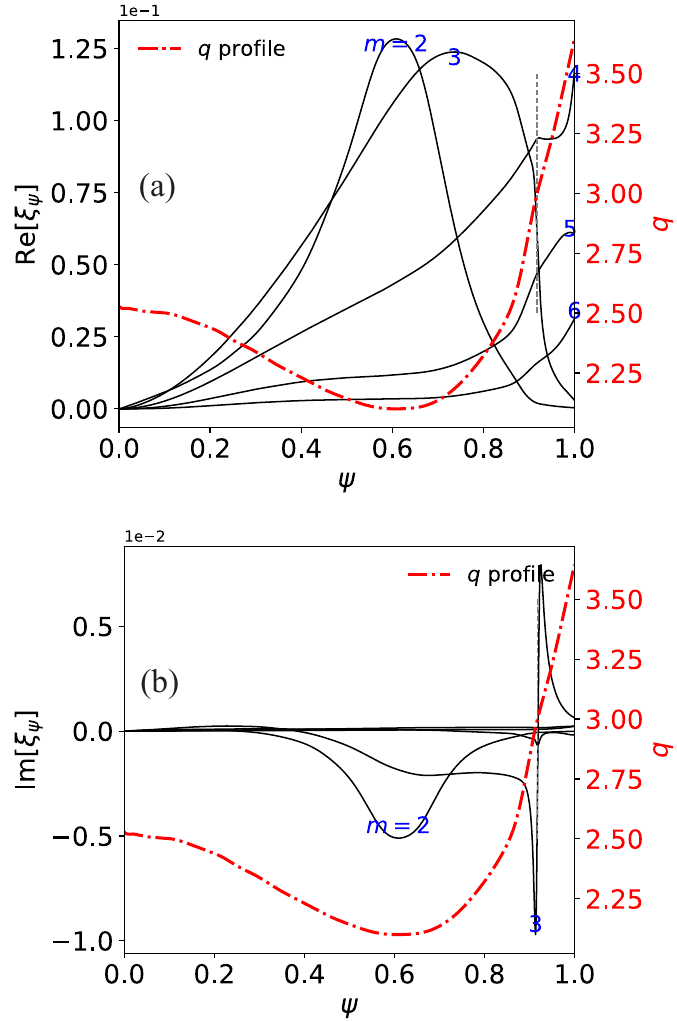}
		\caption{Radial profiles of the real (a) and imaginary (b) parts of the dominant poloidal Fourier components of the radial displacement $\xi_\psi$ for the $n=1$ RWM in the \texttt{circ-rev} equilibrium with $\beta_{\mathrm{N}}=4.91$, $r_w/a=1.2$, and rotation frequency $\Omega/\Omega_{\mathrm{A}}=1\%$. Dominant poloidal mode numbers are indicated.}
		\label{fig:circ-rev-with-rot1e2-mode}
	\end{figure}

\begin{figure}[h]
		\centering
		\includegraphics[width=1.\linewidth]{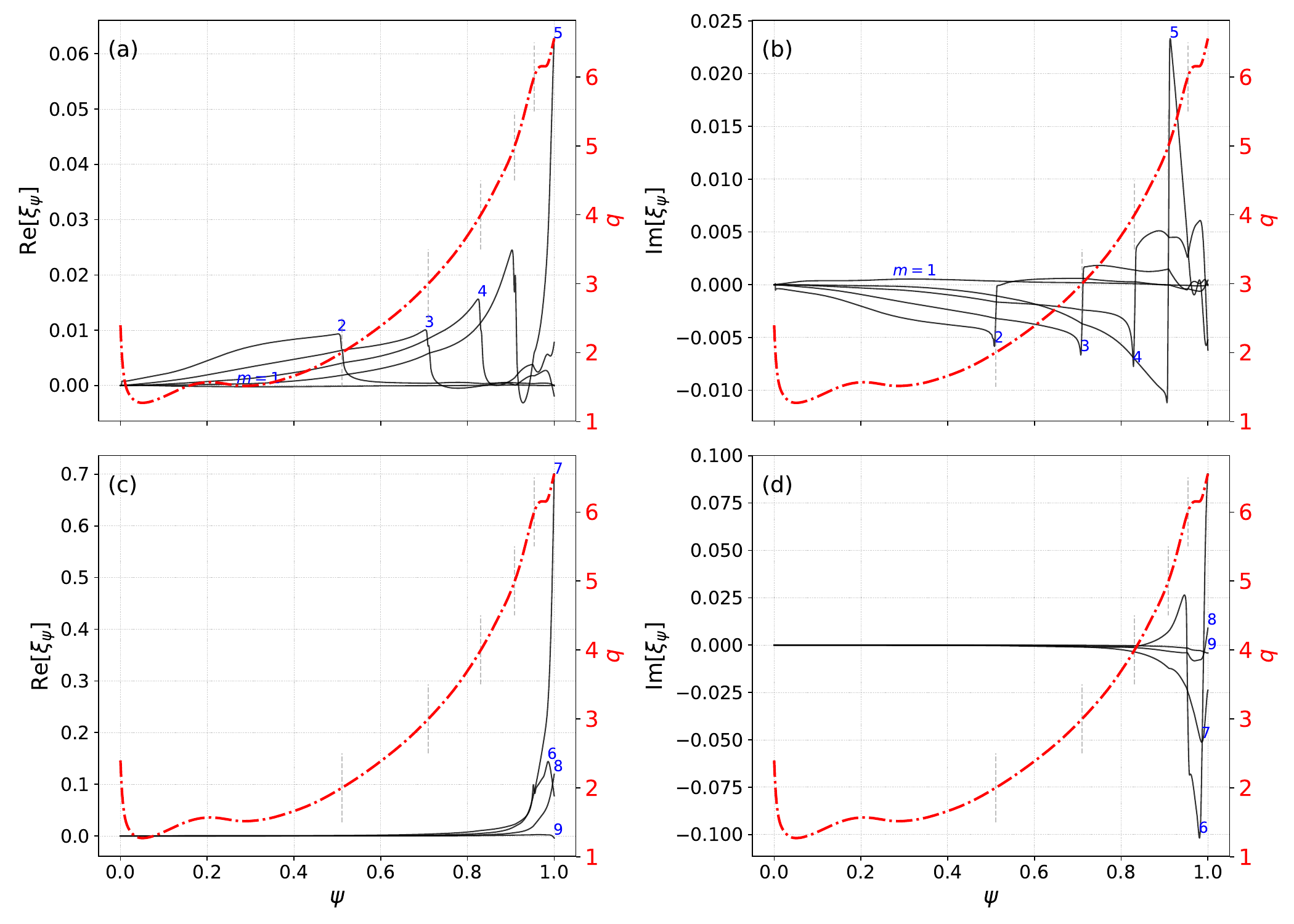}
		\caption{Real (a, c) and imaginary (b, d) parts of the radial displacement $\xi_\psi$ for the $n=1$ RWM in the \texttt{CFETR-1} equilibrium with $\beta_{\mathrm{N}}=1.904$, $r_w/a=1.2$, and rotation frequency $\Omega/\Omega_{\mathrm{A}}=0.2\%$. The poloidal harmonics $m=1-5$ (a, b) and $m=6-9$ (c, d) are indicated accordingly.}
		\label{fig:cfetr-eq1-with-rot2e3-mode}
	\end{figure}

The impact of rotation on the RWM radial profile is illustrated in Figs.~\ref{fig:circ-rev-with-rot1e2-mode} and~\ref{fig:cfetr-eq1-with-rot2e3-mode}. Within the considered range of rotation frequencies ($\Omega/\Omega_A=1\%$ for \texttt{circ-rev}, $\Omega/\Omega_A=0.2\%$ for \texttt{CFETR-1}), complex structures emerge across multiple poloidal harmonics, with the imaginary components becoming more pronounced under rotation, reflecting phase variations and the non-Hermitian character of the eigenvalue problem. These features correspond to the splitting of resonant solutions into complex-conjugate pairs, a hallmark of continuum damping effects in rotating plasmas~\cite{zheng05a,zheng09a,zheng10pop,zheng17a}.

	\subsection{Rotation threshold for complete RWM stabilization}\label{sec4.3}
	
\begin{figure}[h]
		\centering
		\includegraphics[width=0.5\linewidth]{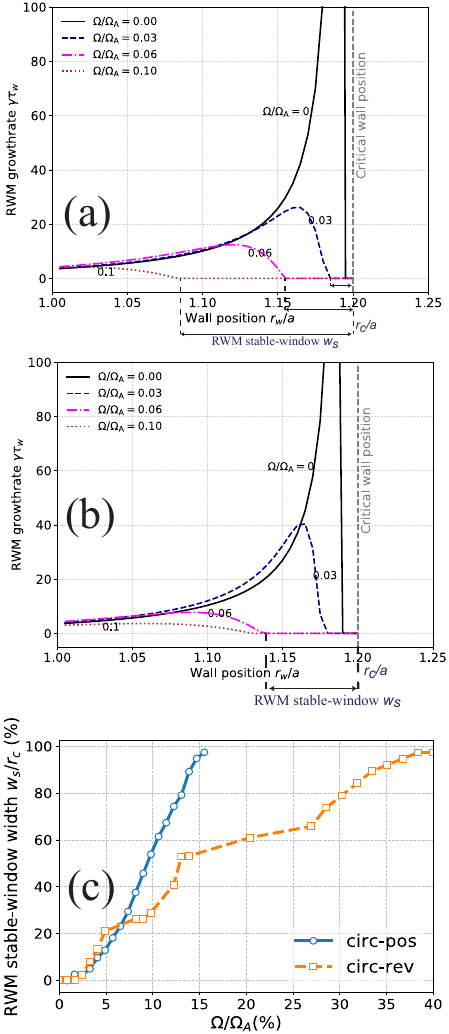}
		\caption{Normalized growth rates ($\gamma\tau_w$) of the $n=1$ RWM as functions of wall radius $r_w/a$ at various rotation frequencies for (a) \texttt{circ-pos} and (b) \texttt{circ-rev} equilibria. The critical wall radius $r_c$ and the stable window width $w_s$ are marked explicitly; (c) normalized width of the stable window ($w_s/r_c$) versus normalized rotation frequency ($\Omega/\Omega_A$) for \texttt{circ-pos} and \texttt{circ-rev} equilibria.}
		\label{fig:circ-RWMgamma-rw-window_revised}
	\end{figure}

\begin{figure}[h]
	    \centering
	    \includegraphics[width=0.9\linewidth]{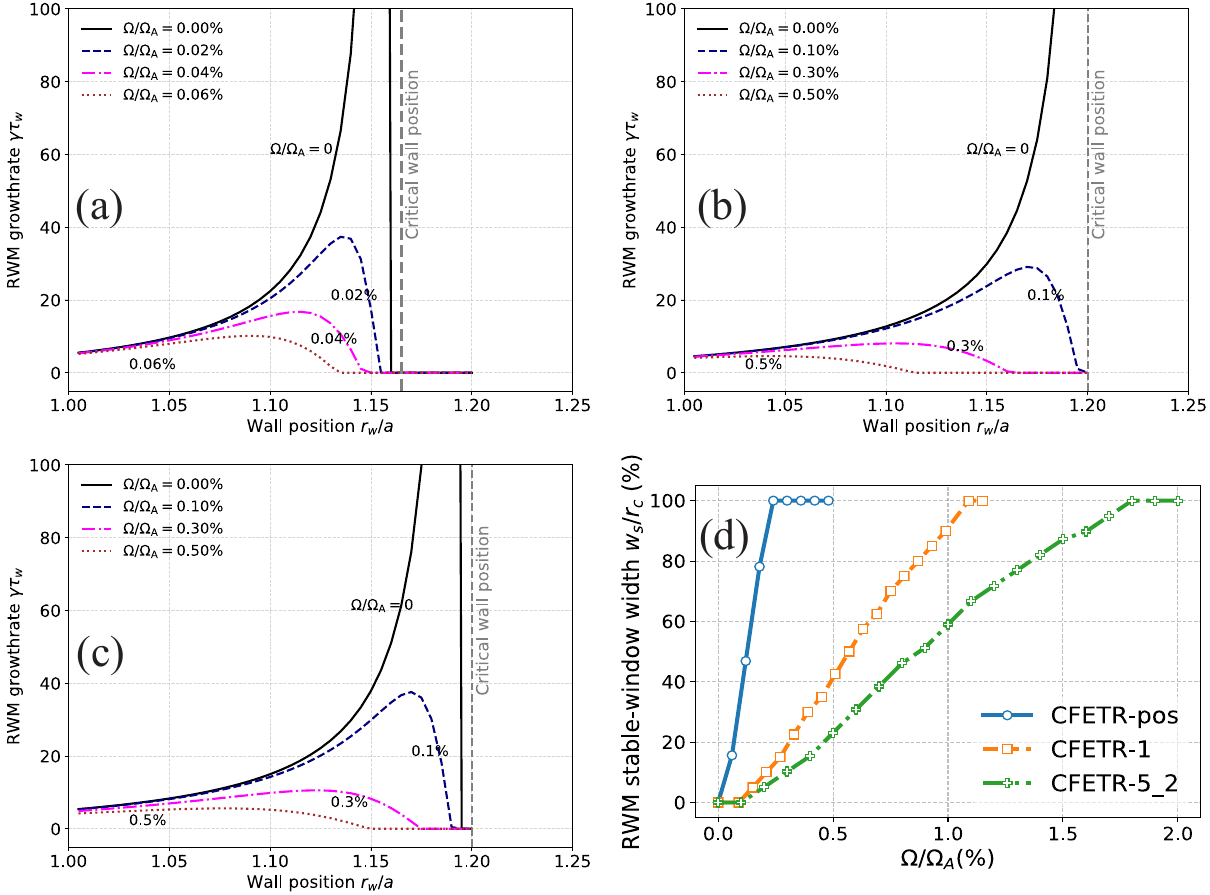}
	    \caption{Normalized growth rates ($\gamma\tau_w$) of the $n=1$ RWM as functions of wall radius $r_w/a$ at various rotation frequencies for CFETR equilibria: (a) \texttt{CFETR-pos}, (b) \texttt{CFETR-1}, and (c) \texttt{CFETR-5\_2}; (d) normalized width of the stable window ($w_s/r_c$) versus normalized rotation frequency ($\Omega/\Omega_A$) for CFETR equilibria.}
	    \label{fig:cfetr-rwmgamma-rw-all_revised}
    \end{figure}

	To achieve robust control of RWM in future fusion devices, it is essential to determine the minimum toroidal rotation frequency required for the full suppression of the instability across all relevant wall locations. Figs.~\ref{fig:circ-RWMgamma-rw-window_revised}(a-b) and~\ref{fig:cfetr-rwmgamma-rw-all_revised}(a-c) compare the $n=1$ RWM stable windows in wall position $r_w/a$ among various rotation frequencies, for circular and CFETR equilibria, respectively. In all cases, the equilibrium pressure is slightly above the corresponding ideal-wall stability limit. In the absence of rotation, the growth rate increases with $r_w$ and diverges as the wall approaches the critical radius $r_c$. When the rotation is included, the growth rate $\gamma\tau_w$ initially increases but then drops rapidly to zero beyond a certain wall radius, indicating the formation of a finite stable window. This stable window, with width $w_s$ explicitly marked in Fig.~\ref{fig:circ-RWMgamma-rw-window_revised}(a), is defined as the radial interval between the critical ideal wall location $r_c$ and the cutoff location where the mode becomes fully stabilized. Stronger rotation results in a wider stable window~\cite{zheng17a}.
	
	To quantitatively assess this stabilization capability, the normalized width of the stable window $w_s/r_c$ is calculated for various magnetic shear configurations (Figs.~\ref{fig:circ-RWMgamma-rw-window_revised}(c), \ref{fig:cfetr-rwmgamma-rw-all_revised}(d)). In the vast majority of the rotation frequency range, the stable window opens more readily with positive magnetic shear and its normalized width increases rapidly as rotation frequency increases, indicating more effective RWM suppression. In contrast, the reversed shear not only delays the onset of stabilization but also leads to consistently narrower windows. For a given rotation frequency, the normalized stable window width $w_s/r_c$ remains significantly smaller in reversed shear cases compared to their positive shear counterparts, demonstrating the reduced efficiency of rotational stabilization in reversed shear configurations. Overall, the magnetic shear configuration plays a critical role in determining the rotational stabilization effectiveness. As the shear gradually changes from positive to negative, the rotation frequency threshold for full RWM stabilization exhibits a monotonic increase. 
	\section{Summary and discussion}\label{sec5}
	
	In this study, the stability of RWMs under various magnetic shear conditions has been systematically investigated using the AEGIS eigenvalue code. Plasma equilibria with circular and shaped cross-sections, including realistic CFETR scenarios, are generated using the CHEASE and EFIT codes, respectively. The primary focus is on evaluating how the influences of uniform toroidal rotation on RWMs may vary from positive to reversed magnetic shear configurations.
	
	Two key aspects have been examined in this study. First, the influence of toroidal rotation on the RWM-unstable $\beta_{\mathrm{N}}$ regime was assessed for the $n=1$ mode. Strongly reversed magnetic shear configurations are found to consistently broaden the unstable $\beta_{\mathrm{N}}$ regime under rotation, in contrast to positive shear configurations, for both circular and CFETR-shaped equilibria. Second, the effects of magnetic shear on the $n=1$ RWM stability window in wall proximity are examined at various rotation frequencies. For most of the rotation frequency range considered, reversed shear equilibria exhibit significantly narrower stable windows than the positive shear cases. In particular, for CFETR equilibria, the critical rotation frequency required for complete RWM stabilization, which is defined as the situation where the stability window in proximity reaches unity, increases progressively with stronger core reversed shear. Thus, achieving comparable RWM stabilization in reversed shear scenarios demands notably higher rotation frequencies than in positive shear equilibria. Overall, these findings illustrate that the reversed shear configurations impose stricter constraints on passive rotational stabilization.
	
	For burning plasmas in advanced tokamaks, outstanding kinetic effects associated with energetic particle and thermal ion resonances may bring in significant stabilizations that could offset the adversary influence of reversed magnetic shear on the rotational stabilization of RWMs. We plan on such studies in future work.
	
	\ack
	This work is supported by the National MCF Energy R\&D Program of China under Grant No.~2019YFE03050004, the U.S. Department of Energy Grant No.~DE-FG02-86ER53218, and the Hubei International Science and Technology Cooperation Project under Grant No.~2022EHB003. The computing work in this paper is supported by the Public Service Platform of High Performance Computing by Network and Computing Center of HUST.
	\section*{References}
	\bibliography{sample-1}
	\clearpage
	\clearpage
	\clearpage
	\clearpage
	\clearpage
	\clearpage
	\clearpage
	\clearpage
	\clearpage
	
	\clearpage
	\clearpage
    \clearpage
\end{document}